# Marginal anisotropy in layered aperiodic Ising systems

P. E. Berche, B. Berche and L. Turban(*)

Laboratoire de Physique du Solide (**) Université Henri Poincaré (Nancy 1), BP 239
F-54506 Vandœuvre lès Nancy Cedex, France

**Abstract.** — Two-dimensional layered aperiodic Ising systems are studied in the extreme anisotropic limit where they correspond to quantum Ising chains in a transverse field. The modulation of the couplings follows an aperiodic sequence generated through substitution. According to Luck's criterion, such a perturbation becomes marginal when the wandering exponent of the sequence vanishes. Three marginal sequences are considered: the period-doubling, paper-folding and three-folding sequences. They correspond to bulk perturbations for which the critical temperature is shifted. The surface magnetization is obtained exactly for the three sequences. The scaling dimensions of the local magnetization on both surfaces, $x_{ms}$ and $\overline{x}_{ms}$, vary continuously with the modulation factor. The low-energy excitations of the quantum chains are found to scale as $L^z$ with the size $L$ of the system. This is the behaviour expected for a strongly anisotropic system, where $z$ is the ratio of the exponents of the correlation lengths in the two directions. The anisotropy exponent $z$ is here simply equal to $x_{ms}+\overline{x}_{ms}$. The anisotropic scaling behaviour is verified numerically for other surface and bulk critical properties as well.

**Résumé.** — On étudie des systèmes d'Ising apériodiques en couches à deux dimensions dans la limite anisotrope extrême où ils correspondent à des chaînes quantiques d'Ising en champ transverse. La modulation des interactions est engendrée par une suite apériodique obtenue par substitution. D'après le critère de Luck, une telle perturbation devient marginale lorsque l'exposant de "divagation" associé à la suite s'annule. Trois suites marginales sont examinées : "doublement de période", "pliage de papier" et "pliage ternaire". Elles correspondent à des perturbations de volume entrainant un changement de température critique. Des expressions exactes de l'aimantation de surface sont obtenues pour les trois suites. Les dimensions anormales de l'aimantation locale sur les deux surfaces, $x_{ms}$ et $\overline{x}_{ms}$, varient continûment avec le facteur de modulation. Les excitations de basse énergie des chaînes quantiques varient en $L^z$ avec la taille $L$ du système. C'est là le comportement attendu pour un système fortement anisotrope où $z$ est le rapport des exposants associés aux longueurs de corrélation dans les deux directions. L'exposant d'anisotropie $z$ s'exprime simplement comme la somme des exposants magnétiques $x_{ms} + \overline{x}_{ms}$. Le comportement d'échelle anisotrope est vérifié numériquement pour d'autres propriétés critiques de surface et de volume.

---

(*) Author for correspondence (e-mail: turban@lps.u-nancy.fr)
(**) Unité de Recherche Associée au CNRS (URA 155)

## 1. Introduction.

The discovery of quasicrystals [1] has opened a new area of research which has been quite active in the last years (see [2–6] for reviews). In the field of critical phenomena, quasiperiodic or aperiodic systems are quite interesting since they offer the possibility to interpolate between periodic and random systems.

The Ising model [7–9], the percolation problem [10, 11] and the statistics of self–avoiding walks [12] were found to display the same critical behaviour on the two-dimensional ($2d$) Penrose lattice as on a periodic one. Universal behaviour was also obtained in $3d$ [13]. On the other hand, an aperiodic modulation of the couplings was shown to influence interface roughening in $2d$ [14, 15]: a continuously varying roughness exponent was obtained with the Fibonacci sequence.

Layered systems provide another type of aperiodic system which is now accessible experimentally in three dimensions, *via* artificially grown multi-layers. Long after the pioneering work of McCoy *et al* [16–19] on the $2d$ randomly layered Ising model, some results have been obtained with an aperiodic modulation of the interlayer couplings [20–25]. In most cases the specific heat displays the Onsager logarithmic singularity but some aperiodic sequences have also been found for which the singularity is washed out [26], like in the randomly layered system.

The situation remained unclear until Luck recently proposed a relevance-irrelevance criterion [27] adapted to aperiodic perturbations. As in the Harris criterion for random systems [28], the strength of the fluctuations of the couplings, on a scale given by the correlation length, is of primary importance for the critical behaviour. Thus an aperiodic perturbation can be relevant, marginal or irrelevant, depending on the sign of a crossover exponent involving the correlation length exponent of the unperturbed system $\nu$ and the wandering exponent $\omega$ which governs the fluctuations of the aperiodic sequence [29, 30]. The criterion has been extended to the case of strongly anisotropic systems with uniaxial aperiodicity [31], explaining the interface roughening results, and later generalized to $d$-dimensional aperiodic systems [32].

Some exact results for the surface magnetization of the $2d$ layered Ising model in the extreme anisotropic limit [33] have been obtained with irrelevant, marginal and relevant aperiodic perturbations [34–37]. The critical behaviour is as expected on the basis of the Luck criterion: one obtains stretched exponentials or first-order surface transition when the perturbation is relevant and power laws with continuously varying exponents in the marginal case. Conformal aspects have been also discussed [38].

In the present work, we continue our study of marginal sequences, through a detailed analysis of the strongly anisotropic scaling behaviour which has been recently found in the layered Ising model [39]. In section 2, we recall some generalities about the layered Ising model in the extreme anisotropic limit and aperiodic sequences generated through substitution. The surface and bulk properties of the layered Ising model with period-doubling (PD), paper-folding (PF) and three-folding (TF) modulations are presented and analysed in the light of anisotropic scaling in section 3. Some properties of the PF and TF sequences are discussed in two appendices.

## 2. Layered Ising model and substitution sequences.

**2.**1 APERIODIC ISING QUANTUM CHAIN. — We consider a $2d$ layered Ising model with constant interaction $K_1$ along the layers and modulated interactions $K_2(k)$ between successive layers, $k$ and $k+1$ as shown in figure 1. The extreme anisotropic limit corresponds to $K_1 \to \infty$, $K_2 \to 0$, while keeping the ratio $\lambda_k = K_2(k)/K_1^*$ fixed, where $K_1^* = -\frac{1}{2}\ln\tanh K_1$ is the dual coupling. In this limit, the row-to-row transfer operator $\mathcal{T} = \exp[-2K_1^*\mathcal{H}]$ involves the Hamiltonian of a

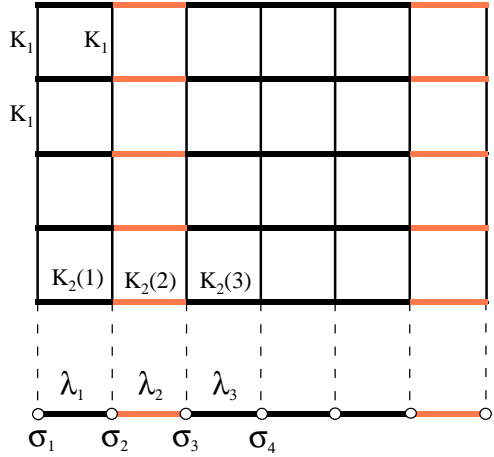

Fig. 1. — Layered aperiodic Ising model in the extreme anisotropic limit (top) and the corresponding quantum chain (bottom)

quantum Ising chain in a transverse field:

$$\mathcal{H} = -\frac{1}{2}\sum_k \left(\sigma_k^z + \lambda_k \sigma_k^x \sigma_{k+1}^x\right) ,\tag{1}$$

where the $\sigma$'s are Pauli spin operators and $\lambda_k$ is the inhomogeneous coupling defined above.

Using the Jordan-Wigner transformation [40] and a canonical transformation to the diagonal fermion operators $\eta_\alpha$, the quantum chain Hamiltonian can be put into diagonal form [41]:

$$\mathcal{H} = \sum_\alpha \epsilon_\alpha \left(\eta_\alpha^\dagger \eta_\alpha - \frac{1}{2}\right) .\tag{2}$$

The fermion excitations $\epsilon_\alpha$ follow from the solution of the system

$$\begin{aligned}\epsilon_\alpha \Psi_\alpha(k) &= -\Phi_\alpha(k) - \lambda_k \Phi_\alpha(k+1) \\ \epsilon_\alpha \Phi_\alpha(k) &= -\lambda_{k-1}\Psi_\alpha(k-1) - \Psi_\alpha(k)\end{aligned}\tag{3}$$

which may be rewritten as a single eigenvalue equation, either for $\underline{\Phi}$ or $\underline{\Psi}$. The components $\Phi_\alpha(k)$ and $\Psi_\alpha(k)$ satisfy appropriate boundary conditions and the eigenvectors are assumed to be normalized.

With an aperiodic modulation, one may write the couplings as:

$$\lambda_k = \lambda r^{f_k} ,\tag{4}$$

where $f_k$, which may be 0 or 1, is determined by the aperiodic sequence. Let

$$n_L = \sum_{k=1}^{L} f_k \tag{5}$$

be the number of modified couplings on a chain with length $L$; their asymptotic density is $\rho_\infty = \lim_{L\to\infty} n_L/L$. The critical coupling $\lambda_c$ is such that [42] $\lim_{L\to\infty}(1/L)\sum_{k=1}^L \ln \lambda_k = 0$, which, using equation (4), gives:

$$\lambda_c = r^{-\rho_\infty} .\tag{6}$$

**2.2 SUBSTITUTION SEQUENCES AND LUCK'S CRITERION.** — The sequences considered below are generated *via* substitutions, either on digits or on pairs of digits.

The PD sequence [43] follows from the substitutions $\mathcal{S}(0) = 1\ 1$ and $\mathcal{S}(1) = 1\ 0$, which, starting on 1, give successively:

$$
\begin{array}{ll}
n = 1 & 1 \\
n = 2 & 1\ 0 \\
n = 3 & 1\ 0\ 1\ 1 \\
n = 4 & 1\ 0\ 1\ 1\ 1\ 0\ 1\ 0
\end{array}
\tag{7}
$$

Details about the properties of the period-doubling sequence can be found in [27, 34, 36].

Two other sequences will be considered in the following: the PF sequence [44] generated through substitutions on pairs of digits, which is studied in appendix A, and the TF sequence [45] studied in Appendix B.

Most of the properties of a sequence can be deduced from its substitution matrix [29, 30] with entries $M_{ij}$ giving the numbers $n_i^{\mathcal{S}(j)}$ of digits (or pairs) of type $i$ in $\mathcal{S}(j)$. In the case of the PD sequence, one obtains:

$$
\underline{\underline{M}} = \begin{pmatrix} n_0^{\mathcal{S}(0)} & n_0^{\mathcal{S}(1)} \\ n_1^{\mathcal{S}(0)} & n_1^{\mathcal{S}(1)} \end{pmatrix} = \begin{pmatrix} 0 & 1 \\ 2 & 1 \end{pmatrix}.
\tag{8}
$$

One may verify that the entries in $\underline{\underline{M}}^n$ give the same numbers of digits (or pairs) in the sequence after $n$ iterations. Let $\underline{V}_\nu$ be the right eigenvectors of the substitution matrix and $\Lambda_\nu$ the corresponding eigenvalues. The asymptotic density of digits (or pairs) of type $i$ are related to the components of the eigenvector corresponding to the leading eigenvalue $\Lambda_1$ through

$$
\rho_{\infty,i} = \frac{V_1(i)}{\sum_j V_1(j)},
\tag{9}
$$

allowing the calculation of the critical coupling *via* equation (6). The length of the sequence after $n$ iterations is related to the leading eigenvalue through $L \sim \Lambda_1^n$ so that $\Lambda_1 > 1$. The cumulated deviation $\Delta(L)$ from the average $\overline{\lambda}$ at the length scale $L$ reached after $n$ iterations may be expressed as

$$
\Delta(L) = \sum_{k=1}^{L} (\lambda_k - \overline{\lambda}) = \lambda(r-1)(n_L - L\rho_\infty) \sim \delta |\Lambda_2|^n \sim \delta L^\omega
\tag{10}
$$

where $\delta = \lambda(r-1)$ is the amplitude of the modulation and $\omega$ is the wandering exponent given by:

$$
\omega = \frac{\ln |\Lambda_2|}{\ln \Lambda_1}.
\tag{11}
$$

Thus the mean shift of the coupling strength at the length scale $L$ is proportional to $L^{\omega-1}$.

Near the critical point of the pure system, the relevant length is the correlation length $\xi \sim t^{-\nu}$ and the aperiodicity induces a shift in the critical temperature $\overline{\delta t} \sim \xi^{\omega-1} \sim t^{-\nu(\omega-1)}$ to be compared with the deviation from the critical point $t$ [27]. One obtains a ratio

$$
\frac{\overline{\delta t}}{t} \sim t^{-\phi}, \qquad \phi = 1 + \nu(\omega - 1),
\tag{12}
$$

which is divergent when $\phi > 0$, which corresponds to a relevant perturbation. When $\phi < 0$, the perturbation, which is washed out when the critical point is approached, is irrelevant. When $\phi = 0$, the perturbation is marginal and may lead to a non-universal behaviour. The same conclusion can be reached by looking for the scaling dimension of the modulation amplitude $\delta$, which is just $\phi/\nu$ [31,34–36].

## 3. Surface and bulk critical behaviour.

We first consider the aperiodic Ising quantum chain with free boundary conditions, i.e. with $\Psi_\alpha(0) = \Phi_\alpha(L+1) = 0$ for a chain with $L$ spins.

**3.**1 SURFACE MAGNETIZATION AND SURFACE ENERGY. — The surface magnetization $m_s$ follows from the asymptotic behaviour when $\tau \to \infty$ of the surface spin-spin correlation function $G_\sigma^s(\tau) = \langle \sigma_1^x(0) \sigma_1^x(\tau) \rangle$ which gives $m_s^2$. In this expression, $\tau$ is an imaginary time and the operators are written in the Heisenberg picture, $\sigma_1^x(\tau) = \exp(\tau \mathcal{H}) \sigma_1^x \exp(-\tau \mathcal{H})$.

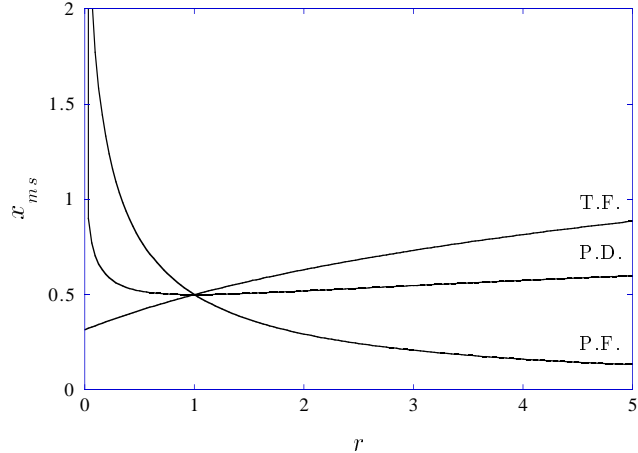

Fig. 2. — Non-universal scaling dimension $x_{ms}$ of the surface magnetization as a function of the modulation factor $r$ for the period-doubling (PD), paper-folding (PF) and three-folding (TF) sequences.

Working in the basis which diagonalizes the Hamiltonian (1), one obtains:

$$G_\sigma^s(\tau) = \sum_{i>0} |\langle i|\sigma_1^x|0\rangle|^2 e^{-\tau(E_i - E_0)} \qquad (13)$$

where $E_0$ is the ground-state energy. Rewriting the matrix element $\langle i|\sigma^x|0\rangle$ and the excited state $|i\rangle$ in terms of diagonal fermions leads to:

$$G_\sigma^s(\tau) = \sum_\alpha \Phi_\alpha^2(1) e^{-\tau \epsilon_\alpha} . \qquad (14)$$

One may notice that one-fermion excited states alone contribute to the sum.

In the ordered phase, $\lambda > \lambda_c$, the lowest excited state $|\sigma\rangle = \eta_1^\dagger|0\rangle$ becomes degenerate with the ground state and gives the only nonvanishing contribution to $m_s^2$ when $\tau \to \infty$ in (14). It follows that $m_s$ is given by the first component $\Phi_1(1)$ of the normalized eigenvector associated with the lowest excitation $\epsilon_1$ [46].

Table I. — *Surface energy exponent for the three marginal sequences studied as function of the modulation factor $r$. For each sequence $x_{es}$ is the extrapolated numerical value whereas the next column gives the conjectured analytical result. The figure in brackets gives the uncertainty on the last digit.*

| | period-doubling | | paper-folding | | three-folding | |
|---|---|---|---|---|---|---|
| $r$ | $x_{es}$ | $3x_{ms} + \overline{x}_{ms}$ | $x_{es}$ | $3x_{ms} + \overline{x}_{ms}$ | $x_{es}$ | $3x_{ms} + \overline{x}_{ms}$ |
| 0.2 | 2.396 71(1) | 2.396 711 | 4.009(1) | 4.008 961 | 1.962(3) | 1.962 149 |
| 0.5 | 2.076 34(4) | 2.076 341 | 2.669 92(5) | 2.669 925 | 1.882 1(2) | 1.881 995 |
| 1   | 2.000 0(4)  | 2         | 2.000 0(3)  | 2         | 2.000 1(2) | 2 |
| 1.5 | 2.026 3(1)  | 2.026 274 | 1.766 41(8) | 1.766 412 | 2.156 873(3) | 2.156 866 |
| 2.0 | 2.076 4(3)  | 2.076 341 | 1.669 9(3)  | 1.669 925 | 2.309 81(1) | 2.309 811 |
| 2.5 | 2.132 4(8)  | 2.132 544 | 1.631 83(6) | 1.631 818 | 2.452 048(9) | 2.452 048 |
| 3.0 | 2.189 29(9) | 2.189 298 | 1.622 6(6)  | 1.622 556 | 2.583 09(7) | 2.583 082 |
| 3.5 | 2.244 59(7) | 2.244 588 | 1.628 84(6) | 1.628 818 | 2.703 9(5)  | 2.703 831 |
| 4.0 | 2.297 6(3)  | 2.297 688 | 1.643 9(7)  | 1.643 856 | 2.815 3(8)  | 2.815 465 |
| 4.5 | 2.348 3(1)  | 2.348 392 | 1.664(2)    | 1.663 976 | 2.919 1(4)  | 2.919 098 |
| 5.0 | 2.396(1)    | 2.396 711 | 1.687(7)    | 1.687 033 | 3.016(3)    | 3.015 708 |

Since $\epsilon_1$ vanishes in the ordered phase, the first relation in (3) provides a recursion for the components of $\underline{\Phi}_1$. After normalization one obtains the surface magnetization of the semi-infinite system as [47]:

$$m_s = \left(1 + \sum_{j=1}^{\infty} \prod_{k=1}^{j} \lambda_k^{-2}\right)^{-1/2}. \tag{15}$$

With an aperiodic system, making use of equations (4) and (5), it may be rewritten as:

$$m_s = [S(\lambda, r)]^{-1/2}, \qquad S(\lambda, r) = \sum_{j=0}^{\infty} \lambda^{-2j} r^{-2n_j}, \qquad n_0 = 0. \tag{16}$$

For the three sequences one obtains a second order surface transition with continuously varying exponents $\beta_s(r)$ (see appendix A for the PF sequence, appendix B for the TF sequence and reference [34] for the PD sequence). This nonuniversal behaviour is linked to the value of $\omega = 0$ which, together with $\nu = 1$ for the 2$d$ Ising model, leads to a marginal behaviour for the three sequences. One may notice that $\nu$ has to keep its unperturbed value in order to have varying exponents. Otherwise the marginality condition $\phi = 0$ in equation (12) would no longer be satisfied in the perturbed system. Thus, as indicated in the appendices, $\beta_s$ is equal to the scaling dimension of the surface magnetization $x_{ms}$ with:

$$x_{ms} = \frac{\ln(r^{1/3} + r^{-1/3})}{2 \ln 2} \quad \text{(PD)}, \qquad x_{ms} = \frac{\ln(1 + r^{-1})}{2 \ln 2} \quad \text{(PF)}, \qquad x_{ms} = \frac{\ln(2 + r)}{2 \ln 3} \quad \text{(TF)}, \tag{17}$$

on the left surface. The variations with $r$ are shown in figure 2.

These exponents are the extreme anisotropic limits of those obtained analytically on classical 2$d$ systems with a finite value of the anisotropy ratio $K_1/K_2$ [48]. Taking the quantum chain limit does not change qualitatively the critical behaviour.

The magnetic scaling dimension $\overline{x}_{ms}$ on the right surface is obtained by exchanging perturbed and unperturbed couplings, so that:

$$\overline{x}_{ms} = x_{ms} \quad \text{(PD)}, \qquad \overline{x}_{ms} = \frac{\ln(1+r)}{2\ln 2} \quad \text{(PF)}, \qquad \overline{x}_{ms} = \frac{\ln(2+r^{-1})}{2\ln 3} \quad \text{(TF)}. \tag{18}$$

The two dimensions are the same for the PD sequence which, apart the last digit, is symmetric.

Let us now consider the surface energy $e_s$, with scaling dimension $x_{es}$, which is given by the ground-state expectation value of $\sigma_1^z$, i.e. $e_s = \langle 0|\sigma_1^z|0\rangle$. Since $e_s$ contains a non-singular part, it is generally difficult to extract its scaling behaviour. One may consider instead the connected surface energy-energy correlation function $G_\varepsilon^s(\tau) = \langle \sigma_1^z(0)\sigma_1^z(\tau)\rangle - e_s^2$ with scaling dimension $2x_{es}$. Working in the diagonal basis, one obtains:

$$\begin{aligned} G_\varepsilon^s(\tau) &= \sum_{i>0} |\langle i|\sigma_1^z|0\rangle|^2 \, e^{-\tau(E_i-E_0)} \\ &= \sum_\alpha \sum_{\beta>\alpha} [(\epsilon_\beta - \epsilon_\alpha)\, \Phi_\alpha(1)\, \Phi_\beta(1)]^2 \, e^{-\tau(\epsilon_\alpha+\epsilon_\beta)} \end{aligned} \tag{19}$$

Only pairs of excitations are involved in the second expression. The reason for this is that $\sigma_1^z$ only couples the ground state to excited states with two fermions. The first non-vanishing matrix element of $\sigma_1^z$ scales like the surface energy. It involves the ground state and the lowest two-fermions excited state $|\varepsilon\rangle = \eta_1^\dagger \eta_2^\dagger |0\rangle$ and, according to (19), may be written as:

$$|\langle \varepsilon|\sigma_1^z|0\rangle| = (\epsilon_2 - \epsilon_1)\, \Phi_1(1)\, \Phi_2(1)\,. \tag{20}$$

At the critical point, on a chain with length $L$, this quantity behaves as $L^{-x_{es}}$. The exponents given in table I were deduced from sequence extrapolations using the BST algorithm [49] with chain sizes $L = 2^2, 2^4, \cdots, 2^{16}$ for the PD and PF sequences and $L = 3, 3^2, \cdots, 3^{10}$ for the TF sequence. The extrapolated values are consistent with the scaling relation

$$x_{es} = 3x_{ms} + \overline{x}_{ms}\,. \tag{21}$$

For the right surface, $x$ has to be changed into $\overline{x}$.

**3.2 LOW-LYING FERMION EXCITATIONS.** — Assuming that the surface component of the second eigenvector in equation (20) has the same scaling behaviour as the first one, i.e. that $\Phi_1(1)\Phi_2(1) \sim L^{-2x_{ms}}$, one is led to conjecture that low-lying excitations scale like $L^{-z}$ with, according to (21)

$$z = x_{ms} + \overline{x}_{ms}\,. \tag{22}$$

This has been verified numerically for the six lowest excitations. The results from sequence extrapolations, with the same sizes as in the last section, are given in table II for the first excitation of the PF sequence.

Such a behaviour is consistent with a strong anisotropy, i.e. with different correlation length exponents, $\nu$ in the space direction (perpendicular to the layers in the $2d$ classical system) and $\nu_\parallel = z\nu$ in the time direction (along the layers) [50]. The dependence of the anisotropy exponent $z$ on $r$ is shown in figure 3 for the three sequences.

One expects the following behaviour for the excitations under a length rescaling by a factor $b$:

$$\epsilon_\alpha(t, t_s, L^{-1}) = b^{-z}\epsilon_\alpha(b^{1/\nu}t, b^{y_{ts}}t_s, bL^{-1})\,, \tag{23}$$

Table II. — *Finite-size scaling exponent for the first excitation and its derivatives with respect to $t$ and $t_s$. The extrapolated numerical values are compared to the conjectured analytical results for the paper-folding sequence. The figure in brackets gives the uncertainty on the last digit.*

|  | paper-folding ||||||
|---|---|---|---|---|---|---|
| $r$ | $z$ || $z-1$ || $z+2x_{ms}$ ||
| 0.5 | 1.084 9(1) | 1.084 963 | 0.085 0(6) | 0.084 963 | 2.669 92(2) | 2.669 925 |
| 1.0 | 1.000 00(3) | 1. | -0.000 1(3) | 0. | 2.000 1(3) | 2. |
| 2.0 | 1.085 1(2) | 1.084 963 | 0.085(1) | 0.084 963 | 1.669 8(3) | 1.669 925 |
| 3.0 | 1.206(4) | 1.207 519 | 0.208(2) | 0.207 519 | 1.622 2(5) | 1.622 556 |
| 4.0 | 1.324(3) | 1.321 928 | 0.321 9(2) | 0.321 928 | 1.644 2(7) | 1.643 856 |
| 5.0 | 1.424 0(3) | 1.423 998 | 0.424 0(3) | 0.423 998 | 1.686 7(9) | 1.687 033 |

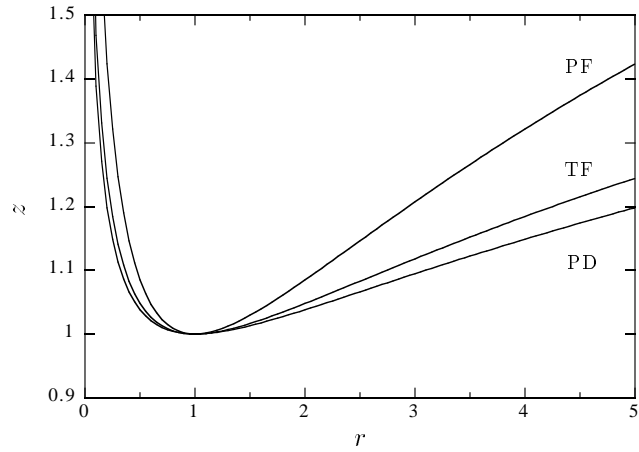

Fig. 3. — Variation of the anisotropy exponent $z$ with the aperiodic modulation factor $r$.

where $t_s$ is the surface thermal scaling field, conjugate to the surface energy, with dimension $y_{ts}$. Due to the *discrete scale invariance* of the sequences, the scaling factor has to take values of the form $m^n$ ($m, n$ integers) where $m = 4, 2, 3$ for the PD, PF and TF sequences, respectively.

At the critical point, as shown in table II, the first derivatives with respect to the bulk and surface scaling fields scale as:

$$\frac{\partial \epsilon_\alpha}{\partial t} \sim L^{1-z} , \qquad \frac{\partial \epsilon_\alpha}{\partial t_s} \sim L^{-z-2x_{ms}} . \qquad (24)$$

These results are consistent with (23) if $\nu = 1$, as found in the last section, and

$$y_{ts} = z - x_{es} = -2x_{ms} , \qquad (25)$$

in agreement with equation (21).

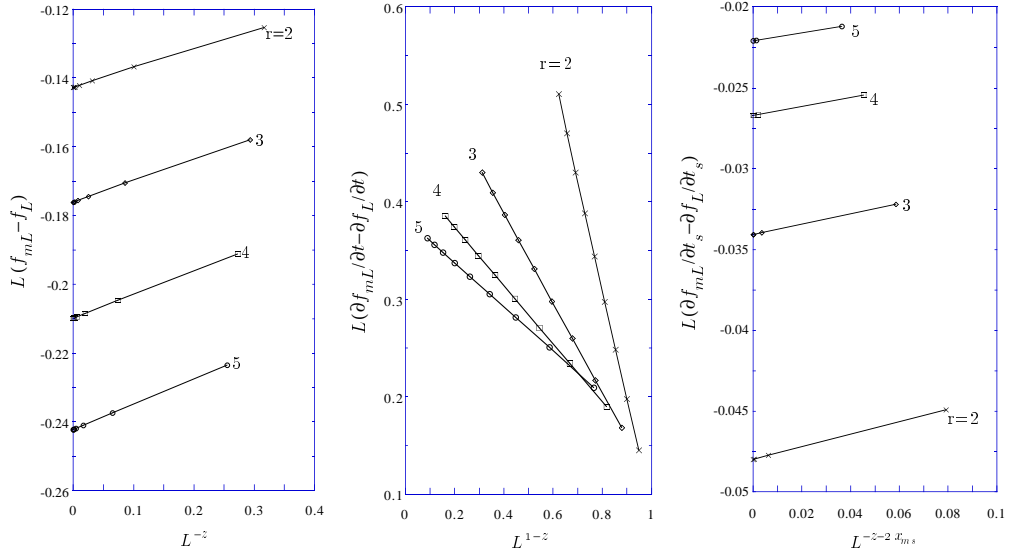

Fig. 4. — Test of the anisotropic finite-size-scaling behaviour of the free energy density for the three-folding sequence with $m = 3$ and chain sizes $L = 3^1$ to $3^{10}$.

Table III. — *Specific heat exponent for the three marginal sequences studied as function of the modulation factor $r$. For each sequence $\alpha$ is the extrapolated numerical value. The next column gives the conjectured analytical value $1 - z$.*

|   | period-doubling | | paper-folding | | three-folding | |
|---|---|---|---|---|---|---|
| $r$ | $\alpha$ | $1-z$ | $\alpha$ | $1-z$ | $\alpha$ | $1-z$ |
| 0.25 | -0.151(5) | -0.148 844 | -0.29(5) | -0.321 928 | -0.181(4) | -0.184 535 |
| 0.5  | -0.039(4) | -0.038 170 | -0.08(1) | -0.084 963 | -0.046(3) | -0.047 952 |
| 0.75 | -0.009(5) | -0.006 623 | -0.011(5) | -0.014 874 | -0.009(2) | -0.008 351 |
| 1.5  | -0.015(3) | -0.013 137 | -0.028(2) | -0.029 447 | -0.015(4) | -0.016 552 |
| 2.0  | -0.045(8) | -0.038 170 | -0.081(4) | -0.084 963 | -0.050(4) | -0.047 952 |
| 2.5  | -0.069(6) | -0.066 272 | -0.152(7) | -0.146 391 | -0.086(6) | -0.082 978 |
| 3.0  | -0.11(3)  | -0.094 649 | -0.207(2) | -0.207 519 | -0.117(8) | -0.118 109 |
| 3.5  | -0.13(1)  | -0.122 294 | -0.26(1)  | -0.266 247 | -0.157(7) | -0.152 102 |
| 4.0  | -0.149(3) | -0.148 844 | -0.31(1)  | -0.321 928 | -0.186(3) | -0.184 535 |
| 4.5  | -0.172(4) | -0.174 196 | -0.36(3)  | -0.374 469 | -0.217(2) | -0.215 310 |
| 5.0  | -0.193(7) | -0.198 356 | -0.45(4)  | -0.423 998 | -0.244(3) | -0.244 464 |

**3.3** FREE ENERGY DENSITY AND SPECIFIC HEAT. — The free energy per row of the system is the ground-state energy of the quantum chain which, according to equation (2), is given by:

$$E_0 = -\frac{1}{2} \sum_\alpha \epsilon_\alpha \ . \tag{26}$$

On a finite system with length $L$, it contains a bulk contribution, proportional to $L$, and a surface contribution. Thus the free energy density can be written as:

$$f(t,t_s,h_s,L^{-1}) = \frac{E_0}{L} = a(t) + \frac{1}{L}b(t,t_s,h_s) + f_b(t,L^{-1}) + \frac{1}{L}f_s(t,t_s,h_s,L^{-1}), \qquad (27)$$

where the two last terms are the singular parts of the bulk and surface contributions, respectively. The regular part has been expanded up to terms of first order in $L^{-1}$ only.

Under rescaling, the singular parts behave as [50]

$$\begin{aligned} f_b(t,L^{-1}) &= b^{-1-z}f_b(b^{1/\nu}t,bL^{-1}), \\ f_s(t,t_s,h_s,L^{-1}) &= b^{-z}f_s(b^{1/\nu}t,b^{y_{ts}}t_s,b^{y_{hs}}h_s,bL^{-1}), \end{aligned} \qquad (28)$$

where $y_{ts}$ is given by (25), $y_{hs} = z - x_{ms} = \overline{x}_{ms}$ and $b = m^n$. Taking a first derivative of the surface free energy with respect to $h_s$ or $t_s$, one recovers the proper scaling behaviour for the surface magnetization or surface energy, respectively.

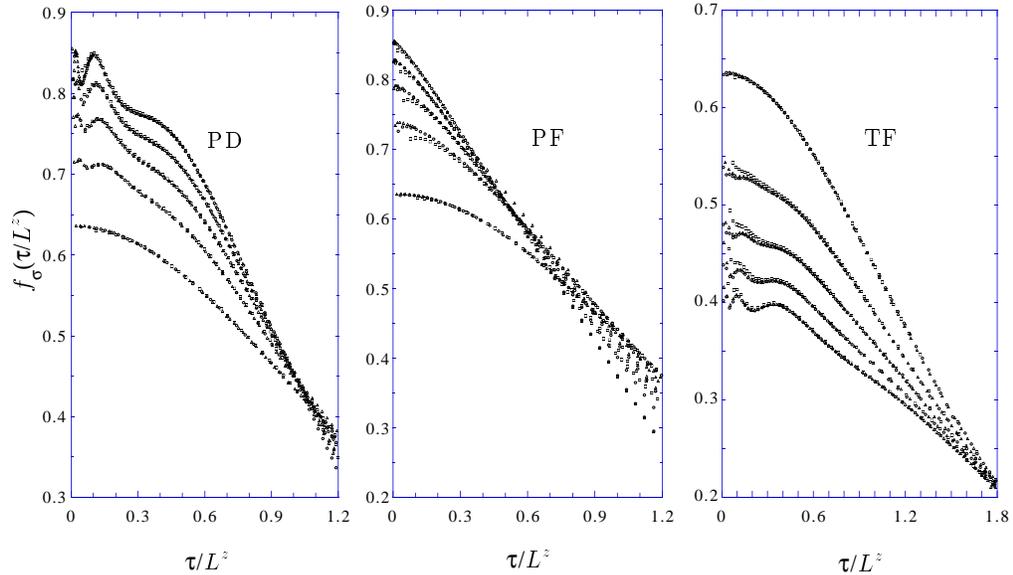

Fig. 5. — Scaling functions for the surface spin-spin correlation function. For the period-doubling (left) and paper-folding (middle) sequences, $r = 5, 4, 3, 2, 1$ from top to bottom and $L = 2^6$ (squares), $2^8$ (triangles) and $2^{10}$ (circles). For the three-folding sequence (right), $r = 1, 2, 3, 4, 5$ from top to bottom and $L = 3^4$ (squares), $3^5$ (triangles) and $3^6$ (circles).

At the critical point, with $f_L = f(0,0,0,L^{-1})$ a linear variation of $L(f_{mL} - f_L)$ with $L^{-z}$ is expected. In the same way, taking the first derivatives with respect to the bulk and surface couplings, one expects linear variations for $L(\partial f_{mL}/\partial t - \partial f_L/\partial t)$ versus $L^{1-z}$ and for $L(\partial f_{mL}/\partial t_s - \partial f_L/\partial t_s)$ versus $L^{-z-2x_{ms}}$. This has been verified numerically for the three sequences. The results for the TF sequence with $m=3$ are shown in figure 4.

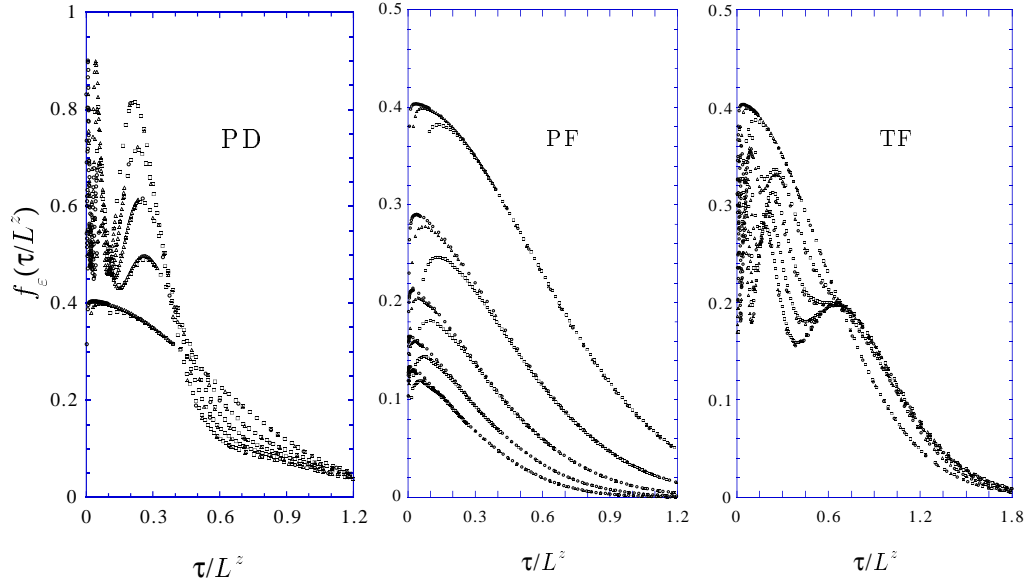

Fig. 6. — Scaling functions for the surface energy-energy correlation function. The values of $r$ and $L$ are the same as in figure 5.

The second derivatives of $f$ with respect to $t$ at the critical point has a leading contribution from its singular part which behaves as $L^{1-z}$, giving the critical exponent of the bulk specific heat

$$\alpha = 1 - z \,. \tag{29}$$

The regular part of the free energy density contributes a constant plus an $L^{-1}$ term which can be neglected as long as $z < 2$. The numerical estimates of $\alpha$, obtained by extrapolating the slopes of $\ln(\partial^2 f_{mL}/\partial t^2 - \partial^2 f_L/\partial t^2)$ versus $\ln L$, are shown in table III with $m=4$ and $L=2^2$ to $2^{16}$ for the PD and PF sequences, $m=3$ and $L=3$ to $3^{10}$ for the TF sequence.

**3**.4 SURFACE CORRELATION FUNCTIONS. — Once the excitation spectrum and the corresponding eigenvectors are known, one may calculate the surface spin-spin and energy-energy correlation functions as given in equations (14) and (19). This has been done at the critical point, on finite chains with size $L$, where the correlation functions scale like:

$$\begin{aligned} G_\sigma^s(\tau, L) &= b^{-2x_{ms}} G_\sigma^s(b^{-z}\tau, b^{-1}L) \,, \\ G_\varepsilon^s(\tau, L) &= b^{-2x_{es}} G_\varepsilon^s(b^{-z}\tau, b^{-1}L) \,, \end{aligned} \tag{30}$$

which gives:

$$\begin{aligned} G_\sigma^s(\tau, L) &= \tau^{-2x_{ms}/z} f_\sigma\left(\frac{\tau}{L^z}\right) = \tau^{-2x_{ms}/z} \left[f_\sigma(0) + \frac{\tau}{L^z} f_\sigma'(0) + \cdots\right] \,, \\ G_\varepsilon^s(\tau, L) &= \tau^{-2x_{es}/z} f_\varepsilon\left(\frac{\tau}{L^z}\right) = \tau^{-2x_{es}/z} \left[f_\varepsilon(0) + \frac{\tau}{L^z} f_\varepsilon'(0) + \cdots\right] \,. \end{aligned} \tag{31}$$

Figures 5 and 6 show the scaling functions for the surface spin-spin and energy-energy correlation functions, respectively. Table IV gives the decay exponents deduced from the slopes of log-log plots using sequence extrapolations for $\tau = 10$ to $100$ and the maximum chain sizes, $L = 2^{10}$ for the PD and PF sequences, $L = 3^6$ for the TF sequence. With the symmetric PD sequence, $x_{ms} = \overline{x}_{ms}$ and the decay exponents remain constant in agreement with equations (21), (22) and (31). They keep their unperturbed values $\eta_\parallel^\sigma = 1$ and $\eta_\parallel^\varepsilon = 4$.

Table IV. — *Decay exponent for the spin-spin ($\eta_\parallel^\sigma$) and energy-energy ($\eta_\parallel^\varepsilon$) surface correlation functions deduced from sequence extrapolations on the slopes of log-log plots at the critical point. Each second column gives the conjectured analytical values.*

|   | period-doubling | | paper-folding | | three-folding | |
|---|---|---|---|---|---|---|
| $r$ | $\eta_\parallel^\sigma$ | $2x_{ms}/z$ | $\eta_\parallel^\sigma$ | $2x_{ms}/z$ | $\eta_\parallel^\sigma$ | $2x_{ms}/z$ |
| 0.5 | 1.04(5) | 1 | 1.459(2) | 1.460 845 | 0.76(5) | 0.795 880 |
| 1 | 0.998(7) | 1 | 0.98(2) | 1 | 0.98(5) | 1 |
| 2.0 | 1.02(3) | 1 | 0.55(2) | 0.539 155 | 1.22(3) | 1.204 120 |
| 3.0 | 1.05(7) | 1 | 0.36(2) | 0.343 711 | 1.30(2) | 1.310 225 |
| 4.0 | 1.04(6) | 1 | 0.249(7) | 0.243 529 | 1.34(4) | 1.376 852 |
| 5.0 | 1.05(5) | 1 | 0.188(7) | 0.184 715 | 1.47(5) | 1.423 298 |
| $r$ | $\eta_\parallel^\varepsilon$ | $2x_{es}/z$ | $\eta_\parallel^\varepsilon$ | $2x_{es}/z$ | $\eta_\parallel^\varepsilon$ | $2x_{es}/z$ |
| 0.5 | 4.01(4) | 4 | 4.92(4) | 4.921 691 | 3.54(5) | 3.591 760 |
| 1 | 3.997(5) | 4 | 4.06(7) | 4 | 4.01(2) | 4 |
| 2.0 | 4.01(2) | 4 | 3.13(6) | 3.078 309 | 4.42(2) | 4.408 240 |
| 3.0 | 4.03(5) | 4 | 2.686(5) | 2.687 422 | 4.613(9) | 4.620 449 |
| 4.0 | 4.02(4) | 4 | 2.47(4) | 2.487 058 | 4.75(2) | 4.753 704 |
| 5.0 | 4.04(6) | 4 | 2.34(5) | 2.369 431 | 4.88(5) | 4.846 596 |

The scaling functions in figures 5 and 6 are clearly oscillating for the PD and TF sequences. Such a behaviour is linked with the discrete scale invariance. Renormalization group arguments show that critical amplitudes may be log-periodic in such systems (see for instance [51]). When the scaling factor is allowed to take values of the form $b = m^n$, the amplitude is a periodic function with period 1 in the variable $\ln\theta/\ln m$ where $\theta$ is some scaling field.

Here, one expects the correlation function to be modulated by an amplitude $A$ which, due to the anisotropic scaling, is a periodic function with period 1 in the variable $\ln\tau/(z\ln m)$. Multiplying both sides of (31) by an oscillating amplitude $A$, it may be transformed on the right into:

$$A\left[\frac{\ln(\tau L^{-z})}{z\ln m}\right] = A\left(\frac{\ln\tau}{z\ln m}\right) \tag{32}$$

when $L = m^n$, thus explaining the good data collapse obtained for the scaling functions.

One may notice that $m$ is generally equal to the leading eigenvalue of the substitution matrix $\Lambda_1$ [30]. Although this is verified for the PF and TF sequences, $m = \Lambda_1^2$ for the PD sequence. It may be verified on (7) that the behaviour of this sequence is different for odd and even values of $n$.

**3.5 BULK MAGNETIZATION.** — On a chain with free boundary conditions, the local magnetization on site $l$ is a matrix element of $\sigma_l^x$ which can be evaluated in the diagonal basis, as in

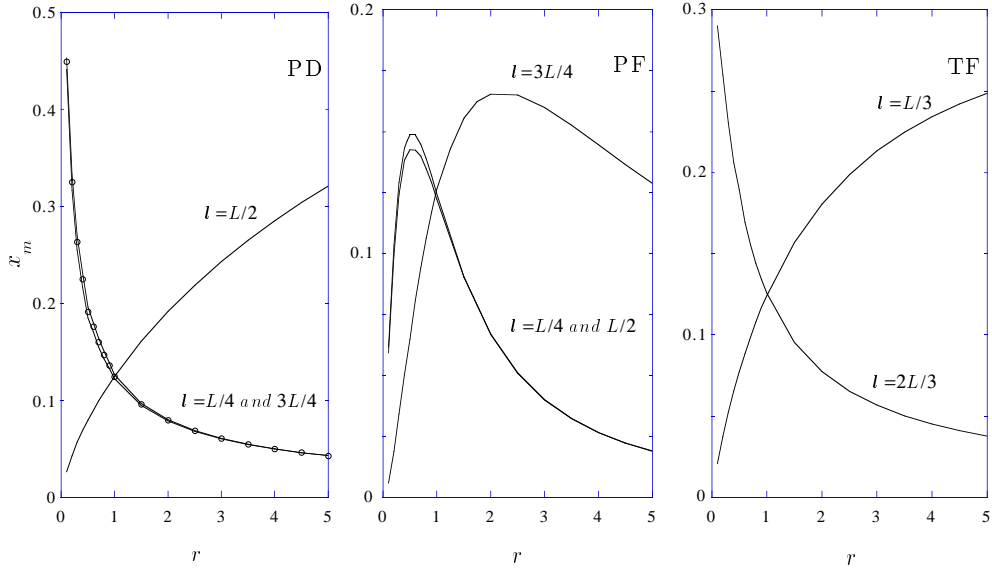

Fig. 7. — Bulk magnetic exponent deduced from finite-size scaling and sequence extrapolations as a function of the modulation amplitude for different values of the ratio $l/L$. The circles for the period-doubling sequence are obtained when the unperturbed coupling at $l=L/2$ is replaced by a perturbed one.

section 3.1, giving:

$$\begin{aligned}
m_l &= \langle\sigma|\sigma_l^x|0\rangle = \langle 0|\eta_1 A_1 B_1 A_2 B_2 \cdots A_{l-1} B_{l-1} A_l|0\rangle\,, \\
A_k &= \sum_\alpha \Phi_\alpha(k)\,(\eta_\alpha^\dagger + \eta_\alpha)\,, \\
B_k &= \sum_\alpha \Psi_\alpha(k)\,(\eta_\alpha^\dagger - \eta_\alpha)\,.
\end{aligned} \qquad (33)$$

The local magnetization $m_l$ can be calculated, like correlation functions in reference [41], by means of Wick's theorem. The ground-state expectation value is given by sums of products of pair expectation values. Since $\langle 0|A_j A_k|0\rangle = \langle 0|B_j B_k|0\rangle = 0$ when $j \neq k$, the local magnetization is obtained as the following determinant:

$$m_l = \begin{vmatrix} H_1 & G_{11} & G_{12} & \cdots & G_{1\,l-1} \\ H_2 & G_{21} & G_{22} & \cdots & G_{2\,l-1} \\ \vdots & \vdots & \vdots & & \vdots \\ H_l & G_{l1} & G_{l2} & \cdots & G_{l\,l-1} \end{vmatrix}, \qquad (34)$$

where:

$$\begin{aligned}
H_j &= \langle 0|\eta_1 A_j|0\rangle = \Phi_1(j)\,, \\
G_{jk} &= \langle 0|B_k A_j|0\rangle = -\sum_\alpha \Phi_\alpha(j)\,\Psi_\alpha(k)\,.
\end{aligned} \qquad (35)$$

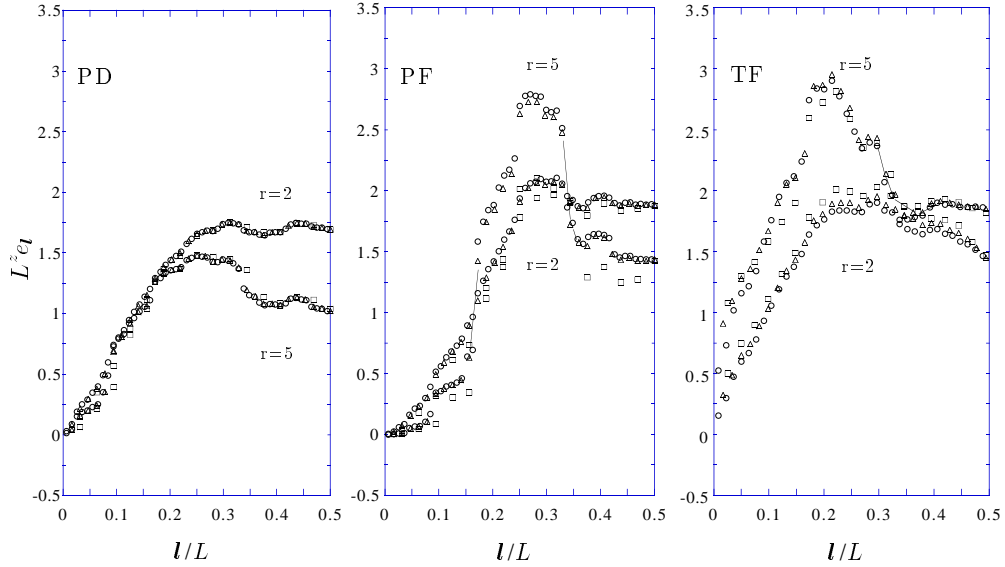

Fig. 8. — Scaled energy density profile for the three sequences with a modulation factor $r=2$ and 5 and chain sizes $L=2^6$ (squares), $2^8$ (triangles), $2^{10}$ (circles) for the period-doubling and paper-folding sequences, $L=3^4$ (squares), $3^5$ (triangles), $3^6$ (circles) for the three-folding sequence.

At the critical point, when $L=m^n$, the magnetization is expected to behave as:

$$m_l(L^{-1}) = b^{-x_m} \, m_{l/b}(bL^{-1}) \qquad (36)$$

so that keeping the ratio $l/L$ constant, $x_m$ governs the size-dependance of the magnetization, $m_l \sim L^{-x_m}$.

The bulk magnetization exponent $x_m$ has been obtained through finite-size scaling and sequence extrapolations on chains with sizes $2^2, 2^4, \cdots, 2^{10}$ for the PD and PF sequences and $3, 3^2, \cdots, 3^6$ for the TF sequence. The results are shown in figure 7. The exponent is continuously varying with $r$ due to the marginal aperiodicity. It depends also on the ratio $l/L$, which means that the scaling relation (36) is only valid locally with the local value of $x_m$.

With the PD sequence, for example, different exponents are obtained at $l=L/2$ on one hand and $l=L/4$ or $3L/4$ on the other hand. This can be understood by looking at the sequence near these sites for long enough chains with sizes of the form $L^{2n}$. In the first case, one finds the following sequence, 1 0 1 1 1 0 1 $\underline{0}$ 1 0 1 1 1 0 1 0, whereas the environment is the same but $\underline{0}$ is replaced by $\underline{1}$ in the other case. This is enough to change the local critical behaviour because a thermal line defect is a marginal perturbation when $\nu = 1$ and $d = 2$ [52]. If one changes the (underlined) unperturbed coupling $\lambda$ at $L/2$ by a perturbed one $\lambda r$, the local exponent becomes the same as for $L/4$ and $3L/4$. The same type of behaviour is obtained with the two other sequences (see figure 7). This inhomogeneous critical behaviour suggests a multifractal statistics as for electronic states on the same structures [53].

**3.6 BULK ENERGY DENSITY AND ITS PROFILE.** — We now consider finite chains with length $L$ and periodic boundary conditions, i.e. with $\Phi_\alpha(L+1) = -\mathcal{Q}\Phi_\alpha(1)$ and $\lambda_0 \Psi_\alpha(0) = -\mathcal{Q}\lambda_L \Psi_\alpha(L)$ where the conserved charge $\mathcal{Q} = \prod_{k=1}^{L} \sigma_k^z$ commutes with the Hamiltonian. Thus one may proceed by replacing the charge operator through one of its eigenvalues, $\mathcal{Q} = \pm 1$ provided the levels with the appropriate charge are retained in each sector (see for instance reference [54]).

Table V. — *Bulk energy exponent for the three marginal sequences studied as function of the modulation factor $r$. For each sequence $x_e$ is the numerical value deduced from sequence extrapolations whereas the next column gives the conjectured analytical value. The figure in brackets gives the uncertainty on the last digit.*

|   | period-doubling | | paper-folding | | three-folding | |
|---|---|---|---|---|---|---|
| $r$ | $x_e$ | $z$ | $x_e$ | $z$ | $x_e$ | $z$ |
| 0.5 | 1.03(4) | 1.038 170 | 1.084 9(7) | 1.084 963 | 1.048(5) | 1.047 952 |
| 1. | 0.999 99(2) | 1 | 0.999 98(3) | 1 | 1.000(1) | 1 |
| 1.5 | 1.02(2) | 1.013 137 | 1.029 4(2) | 1.029 447 | 1.016 5(6) | 1.016 552 |
| 2.0 | 1.06(3) | 1.038 170 | 1.084 95(3) | 1.084 963 | 1.047 8(6) | 1.047 952 |
| 2.5 | 1.09(4) | 1.066 272 | 1.146 38(5) | 1.146 391 | 1.083(1) | 1.082 978 |
| 3.0 | 1.12(5) | 1.094 649 | 1.207 51(7) | 1.207 519 | 1.118(2) | 1.118 109 |
| 3.5 | 1.15(5) | 1.122 294 | 1.266 24(2) | 1.266 247 | 1.152(1) | 1.152 102 |
| 4.0 | 1.18(6) | 1.148 844 | 1.321 92(2) | 1.321 928 | 1.184(1) | 1.184 535 |
| 4.5 | 1.20(6) | 1.174 196 | 1.374 5(6) | 1.374 469 | 1.215(1) | 1.215 310 |
| 5.0 | 1.22(6) | 1.198 356 | 1.424 0(9) | 1.423 998 | 1.244(2) | 1.244 464 |

It has been verified in both sectors that low-lying single-particle excitations scale as $L^z$, like in the system with free boundaries. Furthermore, the free energy density can be analysed as in section 3.3 and is consistent with anisotropic scaling. One may notice that there remains a somewhat unexpected $L^{-1}$ contribution to the free energy density which does not appear in homogeneous systems. It seems that closing the aperiodic sequence onto itself is similar to introducing a defect which contributes a size-independent term to the free energy.

We shall discuss more extensively the scaling of the bulk energy density which, on site $l$, follows from the matrix element

$$e_l = \langle \varepsilon | \sigma_l^z | 0 \rangle = |\Psi_1(l)\Phi_2(l) - \Psi_2(l)\Phi_1(l)|, \tag{37}$$

where the ground-state $|0\rangle$ and the two-fermion eigenstate $|\varepsilon\rangle = \eta_1^\dagger \eta_2^\dagger |0\rangle$ both belong to the even sector. The following behaviour is expected:

$$e_l(t, L^{-1}) = b^{-z} e_{l/b}(b^{1/\nu} t, b L^{-1}) \tag{38}$$

with $b = m^n$ and $\nu = 1$ as already mentioned.

According to (38), at the critical point the scaled energy density profile $L^z e_l$ is a function of $l/L$ only. This is verified in figure 8 for the three sequences. A good data collapse is obtained as long as $r$ is not too large and there is no sign of log-periodic modulation in this case. One may notice that the scaling assumption (38) supposes that the energy density keeps the same scaling dimension $z$ everywhere on the periodic chain, i.e. it does not display the inhomogeneous behaviour observed for the magnetization. Only the amplitude of $L^{-z}$ is varying with $l/L$. The bulk energy exponent deduced from finite-size scaling on the first site is compared to the conjectured analytical result for the three sequences in table V.

### 4. Conclusion.

We have presented the results of a systematic study of the influence of marginal aperiodic perturbations on the critical behaviour of the Ising quantum chain. For the three marginal sequences considered, the surface magnetization has been obtained exactly. The surface magnetic exponents are continuously varying with the modulation amplitude and different values $x_{ms}$ and $\overline{x}_{ms}$ are obtained on the two surfaces when the sequence is asymmetric.

The critical behaviour of the excitation spectrum, the free energy density, the specific heat and the surface correlation functions on systems with free boundary conditions have been studied numerically through finite-size scaling. The results are in agreement with anisotropic scaling with an anisotropy exponent $z$ which is simply given by the sum of the two magnetic surface exponents. The thermal surface and bulk exponents have been also related to $x_{ms}$ and $\overline{x}_{ms}$. As a consequence of the discrete scale invariance, the scaling functions of the surface correlation functions display log-periodic modulations for the PD and TF sequences.

The bulk energy profile and the bulk magnetization obtained on systems with periodic boundary conditions also display anisotropic scaling. The bulk magnetic exponent is continuously varying and may take different values around sites corresponding to simple fractions of the chain length. Although the bulk energy exponent is continuously varying too, this variation is a simple consequence of anisotropic scaling: $x_e = z$, which is varying. The correlation length exponent itself keeps its unperturbed value. This is consistent with the occurence of continuously varying exponents, since the value $\nu = 1$ is necessary to satisfy the marginality condition.

One may notice that the anisotropy exponent $z$ is always $> 1$ for the three sequences studied. As a consequence the specific heat exponent $\alpha = 1 - z < 0$ and the Ising logarithmic singularity is washed out.

The anisotropic critical behaviour associated with these marginal systems breaks the conformal invariance of the critical Ising model. It follows that the gap-exponent relations are no longer valid.

Much remains to be done about these systems. One needs some proofs for the different conjectures about the values of the critical exponents. In particular, an analytical calculation of the anisotropy exponent would be useful. In this paper, we did not examine systematically the log-periodic oscillations of the critical amplitudes. The possibility for a multifractal behaviour of the bulk magnetization has also to be explored. Finally, it would be interesting to consider in detail the case of relevant aperiodic perturbations.

### Acknowledgments.

Collaborations with F. Iglói and M. Henkel during the early stages of this work are gratefully acknowledged. The numerical work was supported by CNIMAT under project No 155C95.

### Appendix A.

### Paper-folding sequence.

The paper-folding sequence [44] results from the recurrent folding of a sheet of paper, right over left, as shown for the first few steps in figure 9. After unfolding, one obtains a succession of up- or down-folds. Associating 1 with an up-fold and 0 with a down-fold leads, after four foldings, to the following sequences:

$$1 \ \underline{1} \ 0 \ \underline{1} \ 1 \ \underline{0} \ 0 \ \underline{1} \ 1 \ \underline{1} \ 0 \ \underline{0} \ 1 \ \underline{0} \ 0 \quad (A.1)$$

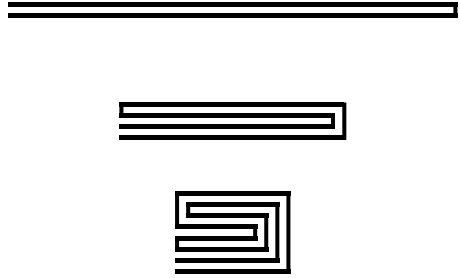

Fig. 9. — Folding $n$ times the right part of a sheet of paper onto the left gives, after unfolding, a succession of up- or down-folds which follow the paper-folding sequence.

The sequence on the right of the central fold is the mirror image of the sequence on the left, with each digit $f_k$ replaced by its complement $1-f_k$. As a consequence, the asymptotic density is $\rho_\infty = \frac{1}{2}$ and equation (6) gives the critical coupling $\lambda_c = r^{-1/2}$.

The same sequence can be generated using the two-digit substitutions $\mathcal{S}(00)=1000$, $\mathcal{S}(01)=1001$, $\mathcal{S}(10)=1100$, $\mathcal{S}(11)=1101$. Starting with 11, the resulting sequence differs from the previous one only through its last supplementary digit, 1, which does not affect the critical behaviour.

According to (11), the leading eigenvalues of the substitution matrix, $\Lambda_1=2$, $\Lambda_2=1$, lead to a vanishing wandering exponent, $\omega=0$, i.e. the corresponding layered perturbation in the $2d$ Ising model is marginal.

Underlined even terms in (A.1) reproduce the sequence whereas odd terms are alternatively 1 or 0 so that we have:

$$f_{2k} = f_k , \qquad f_{2k+1} = \frac{1}{2}\left[1+(-1)^k\right] ,$$
$$n_{2p} = \frac{p}{2} + \frac{3+(-1)^p}{4} + n_p , \qquad n_{2p+1} = \frac{p}{2} + \frac{1-(-1)^p}{4} + n_p . \tag{A.2}$$

The sum in (16) can be rewritten as

$$S(\lambda,r) = \sum_{p=0}^{\infty} \lambda^{-2(2p+1)} r^{-2n_{2p+1}} + \sum_{p=0}^{\infty} \lambda^{-2(2p)} r^{-2n_{2p}} = S_{\text{odd}}(\lambda,r) + S_{\text{even}}(\lambda,r) . \tag{A.3}$$

Separating odd and even values of $p$ in the two sums and using the recursion relations in (A.2), we obtain the following set of coupled functional equations:

$$S_{\text{odd}}(\lambda,r) = \lambda^{-2} r^{-1} S_{\text{odd}}(\lambda^2 r^{1/2}, r) + \lambda^{-2} r^{-2} S_{\text{even}}(\lambda^2 r^{1/2}, r)$$
$$S_{\text{even}}(\lambda,r) = r^{-1} S_{\text{odd}}(\lambda^2 r^{1/2}, r) + S_{\text{even}}(\lambda^2 r^{1/2}, r) . \tag{A.4}$$

These relations can be written in matrix form giving $\underline{V}_0 = \prod_{p=0}^{k-1} \underline{\underline{T}}_p \underline{V}_k$ where:

$$\underline{V}_p = \begin{pmatrix} S_{\text{odd}}(\lambda_p, r) \\ S_{\text{even}}(\lambda_p, r) \end{pmatrix} , \qquad \underline{\underline{T}}_p = \begin{pmatrix} \lambda_p^{-2} r^{-1} & \lambda_p^{-2} r^{-2} \\ r^{-1} & 1 \end{pmatrix} , \qquad \lambda_p = \lambda_c \left(\frac{\lambda}{\lambda_c}\right)^{2^p} . \tag{A.5}$$

In the ordered phase $\lambda > \lambda_c$ and $\lim_{p\to\infty} \underline{V}_p = \begin{pmatrix} 0 \\ 1 \end{pmatrix}$ so that, according to (A.4),

$$S(\lambda, r) = (1\ 1) \left[ \prod_{p=0}^{\infty} \underline{\underline{T}}_p \right] \begin{pmatrix} 0 \\ 1 \end{pmatrix} . \tag{A.6}$$

which, together with equation (16), formally gives the surface magnetization.

The critical behaviour can be obtained using the scaling method of reference [55]. Let $S(z)$ be the series expansion of $S(\lambda, r)$ in powers of $z = (\lambda_c/\lambda)^2$. According to equation (16), near the critical point $S(z)$ behaves as $(1-z)^{-2\beta_s}$. Then it may be shown that the truncated series $S_L(z)$, containing the first $L$ terms in $S(z)$, diverges as $L^{2\beta_s}$ at the critical point $z=1$ when $L \to \infty$. According to finite-size scaling, at the critical point of a system with size $L$, $m_s$ vanishes as $L^{-x_{ms}}$ so that $S_L(1)$ also behaves as $L^{2x_{ms}}$. As a consequence $\beta_s = x_{ms}$ and $\nu = 1$ like in the pure system.

The form of the matrix in (A.5) is such that, keeping the first $n$ factors in the infinite product of equation (A.6), gives

$$S_L(z) = (1\ 1) \left[ \prod_{p=0}^{n-1} \begin{pmatrix} z^{2^p} & \lambda_c^2 z^{2^p} \\ \lambda_c^2 & 1 \end{pmatrix} \right] \begin{pmatrix} 0 \\ 1 \end{pmatrix} . \tag{A.7}$$

with $L = 2^n$. Since at $z = 1$ the matrix becomes $p$-independent, the form of the leading contribution to the truncated series is easily obtained as

$$S_{L=2^n}(1) \sim (2^n)^{2\beta_s} \sim (1 + \lambda_c^2)^n , \tag{A.8}$$

where the last term is the $n$th power of the leading eigenvalue of $\underline{\underline{T}}_p$. Finally, the surface magnetization exponent reads

$$\beta_s = \frac{\ln(1 + r^{-1})}{2 \ln 2} . \tag{A.9}$$

**Appendix B.**

**Three-folding sequence.**

The three-folding sequence [45] is generated through the substitutions $\mathcal{S}(0) = 010$, $\mathcal{S}(1) = 011$. Starting with 0, after four steps one obtains:

$$0\ 1\ \underline{0}\ 0\ 1\ \underline{1}\ 0\ 1\ \underline{0}\ 0\ 1\ \underline{0}\ 0\ 1\ \underline{1}\ 0\ 1\ \underline{1}\ 0\ 1\ \underline{0}\ 0\ 1\ \underline{1}\ 0\ 1\ \underline{0} \tag{B.1}$$

The sequence repeats itself if one keeps every third (underlined) term, which gives:

$$\begin{aligned} f_{3k} &= f_k, & f_{3k+1} &= 0, & f_{3k+2} &= 1. \\ n_{3k} &= n_k + k, & n_{3k+1} &= n_k + k, & n_{3k+2} &= n_k + k + 1. \end{aligned} \tag{B.2}$$

Iterating $n_{3k}$ gives $n_{3p} = \frac{1}{2}(3^p - 1)$ so that, asymptotically, $\rho_\infty = \frac{1}{2}$ and $\lambda_c = r^{-1/2}$. The eigenvalues of the substitution matrix, $\Lambda_1 = 3$, $\Lambda_2 = 1$, lead to the wandering exponent $\omega = 0$, i.e. marginal behaviour in the $2d$ Ising model.

Like in (A.3) the sum $S(\lambda, r)$ may be splitted into three sums which, making use of (B.2), lead to the recursion:

$$\begin{aligned} S(\lambda, r) &= (1 + \lambda^{-2} + r^{-2}\lambda^{-4}) S(\lambda^3 r, r) \\ &= \prod_{p=0}^{\infty} \left( 1 + \lambda_p^{-2} + r^{-2}\lambda_p^{-4} \right) , \end{aligned} \tag{B.3}$$

where $\lambda_p = \lambda_c \, (\lambda/\lambda_c)^{3^p}$. Using the same scaling method and notations as in appendix A, the first $n$ factors in the infinite product contain the first $L=3^n$ terms in the truncated series,

$$S_{L=3^n}(z) = \prod_{p=0}^{n-1} \left[ 1 + rz^{3^p} + (z^2)^{3^p} \right] ,\qquad\text{(B.4)}$$

so that, from its value at the critical point $z=1$, one deduces

$$\beta_s = \frac{\ln(2+r)}{2\ln 3} .\qquad\text{(B.5)}$$